\documentclass[aps,preprint,superscriptaddress]{revtex4}

\usepackage{amsfonts}
\usepackage{eufrak}
\usepackage{amsmath,amssymb,epsfig,natbib,bm}

\def\be{\begin{equation}}
  \def\ee{\end{equation}}
\def\bea{\begin{eqnarray}}
  \def\eea{\end{eqnarray}}

\newcommand{\beqfl}{\bgroup\abovedisplayskip=0pt  \belowdisplayskip=\smallskipamount
                    $$\leqno\bgroup}
\newcommand{\eeqfl}{\egroup$$\egroup}

\begin{document}

\title{Unveiling Hidden Patterns in CMB Anisotropy Maps}

\author{Tuhin Ghosh}
\affiliation{IUCAA, Post Bag 4, Ganeshkhind, Pune-411 007, India}
 \email{tuhin@iucaa.ernet.in;  ahajian@princeton.edu; tarun@iucaa.ernet.in}
\author{Amir Hajian}
\affiliation{Department of Physics, Jadwin Hall, Princeton University,
    Princeton, NJ 08542.}
\affiliation{Department of Astrophysical Sciences, Peyton Hall,
   Princeton University,
    Princeton, NJ 08544.}
\author{Tarun Souradeep}
\affiliation{IUCAA, Post Bag 4, Ganeshkhind, Pune-411 007, India}

\begin{abstract}
Bianchi VII$_h$ models have been recently proposed to explain
potential anomalies in the CMB anisotropy as observed by WMAP. We
investigate the violation of statistical isotropy due to an embedded
Bianchi VII$_h$ templates in the CMB anisotropy maps to determine
whether the existence of a hidden Bianchi template in the WMAP data is
consistent with the previous null detection of the bipolar power
spectrum in the WMAP first year maps. We argue that although
correcting the WMAP maps for the Bianchi template may explain some
features in the WMAP data it may cause other anomalies such as
preferred directions leading to detectable levels of violation of
statistical isotropy in the Bianchi corrected maps. We compute the
bipolar power spectrum for the low density Bianchi VII$_h$ models
embedded in the background CMB anisotropy maps with the power spectrum
that have been shown in recent literature to best fit the first year
WMAP data. By examining statistical isotropy of these maps, we put a
limit of $(\frac{\sigma}{H})_0\le 2.77 \times 10^{-10} (99\% CL)$ on
the shear parameter in Bianchi VII$_h$ models.
\end{abstract}

\maketitle
\section{Introduction}

Friedmann-Robertson-Walker (FRW) models are the simplest models of the
expanding Universe which are spatially homogeneous and isotropic.
When proposed, the principal justification for studying these models
was their mathematical simplicity and tractability rather than
observational evidence. However, we now have observational evidence
from the isotropy of CMB and large scale structure that the universe
on large scales must be very close to that of a FRW model.  However,
there still remains some freedom to choose homogeneous models that are
initially anisotropic but become more isotropic as the time goes on,
and asymptotically tend to a FRW model. Bianchi models provide a
generic description of homogeneous anisotropic cosmologies and they
are classified into 10 equivalence classes
\citep{ellis&maccallum}. Among these models it is reasonable to
consider only those types that encompass FRW models. These are types I
and VII$_0$ in the case of $k\,=\,0$, V and VII$_h$ in the case of
$k\,=\,-1$, and IX in the case of $k\,=\,+1$. Bianchi models which do
not admit FRW solutions become highly anisotropic at late times.  Type
IX models re-collapse after a finite time and hence do not approach
arbitrarily near to isotropy. Also models of type VII$_h$ will not in
general approach isotropy \citep{collins&hawking2}. The most general
Bianchi types that admit FRW models are Bianchi types VII$_h$ and
IX. These two types contain types I, V, VII$_0$ as special
sub-cases. An interesting feature of these models is that they
resemble a universe with a vorticity. It is interesting to determine
bounds on universal rotation from cosmological observations because
the absence of such a rotation is a prediction of most models of the
early universe, in particular, within the paradigm of inflation.

CMB anisotropy is a powerful tool to study the evolution of
vorticity in the universe because models with vorticity have  clear
signatures in the CMB. The vortex patterns that are imprinted by
the unperturbed anisotropic expansion are roughly constructed out of
two parts: production of pure quadrupole variations, or focusing of
the quadrupole pattern  in open models and a spiral
pattern that are the characteristic signatures of VII$_0$ and VII$_h$ models
 respectively \citep{collins&hawking2, doroshke, barrow}.  There are distinct
features in each Bianchi pattern. For example the temperature
anisotropy pattern for Bianchi VII$_h$ universes is of the form \be
\frac{\Delta T^B}{T_0}(\theta,
\phi)\,=\,f_1(\theta)\sin{\phi}+f_2(\theta)\cos{\phi}. \ee The
functions $ f_1(\theta)$ and $f_2(\theta)$ depend on the parameters
of the particular Bianchi model and must be computed numerically
\citep{barrow}. In a pioneering work, analytical arguments were used
to find upper bounds on the amount of shear and vorticity in the
universe today, from the absence of any detected CMB anisotropy
\citep{collins&hawking2}. A detailed numerical analysis of such
models used experimental limits on the dipole and quadrupole to
refine limits on universal rotation \citep{barrow}. After the first
detection of CMB anisotropy by COBE-DMR, Bianchi models were again
studied by fitting the full spiral pattern from models with global
rotation to the 4-year DMR data to constrain the allowed parameters
of a Bianchi model of type VII$_h$ \citep{bunn96, kogut97}. Recently
Bianchi VII$_h$ models were compared to the first-year WMAP data on
large scales  and it was shown that the best fit Bianchi model
corresponds to a highly hyperbolic model ($\Omega_0=0.5$) with a
right-handed vorticity $(\frac{\omega}{H})_0=4.3 \times 10^{-10}$
\citep{jaffe}. They also found that `correcting' the first-year WMAP
data including the ILC map\footnote{The WMAP team warns that the
original ILC map should not be used for detailed tests.}  for the
Bianchi template, makes the reported anomalies in the WMAP data such
as alignment of quadrupole and octupole disappear and also ameliorates 
the problem of 
the low observed value of quadrupole\footnote{Recently they have revised
their results to a slightly different best fit Bianchi VII$_h$ model
with a left handed $(x,\Omega_0)=(0.62,0.15)$ model and a right
handed $(x,\Omega_0)=(0.62,0.5)$ with a larger vorticity
$(\sigma/H)_0=4.33\pm 0.82$ for ILC map \cite{Jaffe:2006fh}.}. In a
more recent work using template fitting  it was found
that although the ``template'' detections are not statistically
significant they do correct  the above anomalies \cite{Land:2005cg}.
On the other hand, it was also found by \cite{jaffe} that deviations
from Gaussianity in the kurtosis of spherical Mexican hat wavelet
coefficients  of the WMAP first year data are eliminated once the
data is corrected for the Bianchi template. In a recent work
 the effect of this Bianchi correction on the
detections of non-Gaussianity in the WMAP data was investigated
\cite{McEwen} and was shown that previous detections of
non-Gaussianity observed in the skewness of spherical wavelet
coefficients  reported in \cite{McEwen2004} are not reduced by the
Bianchi correction and remain at a significant level.
It has been argued in Ref. \cite{Cayon:2006aq}  that increasing
 the scaling of the
template by a factor of 1.2 makes the spot vanish. Although a stronger
pattern may eliminate the non-Gaussian spot, it would tend to make
 the resultant map
more anisotropic. Therefore there is a limit to the strength of
anisotropic patterns that can be hidden in the WMAP data. The above
analysis was redone after the release of 3 year WMAP data
\cite{Hinshaw:2006ia}, by two teams \cite{Bridges:2006mt,
Jaffe:2006sq} and the previous conclusions were confirmed.

In this paper we show that the statistical isotropy violation of a
hidden pattern in the CMB map, such as that of Bianchi universe, can
be quantified in terms of the Bipolar power spectrum. More
specifically, we test the consistency of existence of a hidden Bianchi
template in the WMAP data proposed in recent literature against our
null detection of bipolar power spectrum (BiPS) in the WMAP first year
and three year maps\citep{us_prd, Souradeep:2006dz, Hajian:2006ud}.
The bipolar power spectrum is a measure of statistical isotropy in CMB
anisotropy maps and is zero when statistical isotropy
obtains. Properties of BiPS have been studied in great details in
\cite{us_bigpaper} and \cite{us_polarization}.  We show that although
correcting the WMAP first year maps for the Bianchi template may
explain some features in the WMAP data\footnote{Several methods have
been used to look for possible anomalies in the WMAP data, here is a
list of some of them:
\cite{erik04a,Copi:2003kt,Schwarz:2004gk,Hansen:2004vq,angelwmap,Land:2004bs,Land:2005ad,Land:2005dq,Bielewicz:2004en,Bielewicz:2005zu,Land:2005jq,Copi:2005ff,Land:2005cg,
Naselsky:2004gm,us_prd,Prunet:2004zy,Gluck:2005td,Chen:2005ev,Freeman:2005nx,Stannard:2004yp,Bernui:2005pz,Bernui:2006ft,
SadeghMovahed:2006em, Wiaux:2006zh}.}, this is done at the expense of
introducing some anomalies such as preferred directions and the
violation of statistical isotropy into the Bianchi corrected maps.
This violation is stronger in the case of enhanced Bianchi templates
proposed by \cite{Cayon:2006aq}. By testing statistical isotropy of
Bianchi embedded CMB maps, we put a limit of $(\frac{\sigma}{H})_0 \le
2.77 \times 10^{-10} (99\% CL)$ on Bianchi VII$_h$ models.  Here
$\sigma$ is the shear and $H$ is the Hubble constant. Our result is
marginally consistent with \cite{jaffe} but not with the enhanced
Bianchi templates of \cite{Cayon:2006aq}.

The rest of this paper is organized as the following. Section 2 is a
brief review of Bianchi classification of homogeneous spaces.
Section 3 is dedicated to deriving the CMB patterns of Bianchi
VII$_h$ models. In $\S$~4 statistical isotropy of Bianchi VII$_h$
models are studied and compared to the null bipolar power spectrum
of the WMAP data. Finally in $\S$~5 we draw our conclusion.

\section{Bianchi Models}
Today, the Bianchi classification of homogeneous spaces is based on
a simple scheme for classifying the equivalence classes of
3-dimensional Lie algebras \footnote{All Killing vector fields
 form a Lie algebra of the symmetric
operations on the space, with structure constants $C^a_{bc}$. See \cite{mythesis} for mathematical preliminaries and details.}.
 This scheme uses the irreducible parts
of the structure constant tensor under linear transformations.
Following \cite{ellis&maccallum}, we decompose the spatial
commutation structure constants $C^a_{bc}$ into a tensor, $n^{ab}$,
and a vector, $a_b$ \be C^a_{bc}=\epsilon_{dbc}n^{ad}+\delta^a_c
a_b-\delta^a_b a_c \ee where $\epsilon_{abc}$ is the 3-dimensional
antisymmetric tensor and $n^{ab}$ and $a_b$ are defined as 
\bea
a_b&=& \frac{1}{2} C^a_{ba}
\\ \nonumber n^{ab}&=&\frac{1}{2} C^{(a}_{cd}\epsilon^{b)cd}.
 \eea
The structure constants $C^a_{bc}$ expressed in this way, satisfy
the first Jacobi identity. Second Jacobi identity, shows that $a_b$
must have zero contraction with the symmetric $2$-tensor $n^{ab}$;
\be 
C_{e[b}^aC_{cd]}^e=0 \Longrightarrow n^{ab}a_b=0. 
\ee 
We choose a
convenient basis to diagonalize $n^{ab}$ to attain $n^{ab} =
\mathrm{diag}(n_1, n_2, n_3)$ and to set $a_b = (a, 0, 0)$. The
Jacobi identities are then simply equivalent to $n_1 a = 0$.
Consequently we can define two major classes of structure constants\\
Class A : $a = 0$,\\
Class B : $a\ne0$.\\
One can then classify further by the sign of the eigenvalues of
$n^{ab}$ (signs of $n_1$, $n_2$ and $n_3$).
Parameter $h$ in class B, where $a$
is non-zero, is defined by the scalar constant of proportionality in
the following relation \be
a_ba_c=\frac{h}{2}\epsilon_{bik}\epsilon_{cjl}n^{ij}n^{kl}. \ee
 In the case of diagonal $n^{ab}$, the $h$ factor has a simple form
$h=a^2/(n_2n_3)$.


\section{CMB Patterns in Type VII$_h$ Bianchi Models}
 This is the
most general family of models which includes the $k=-1$ FRW
solutions. It has some of the features of both types V and type
VII$_0$. There is an adjustable parameter $h$ in this model which is
given by the square of structure constants. In an unperturbed FRW
model, the expansion scale factor, $\alpha$, and $\beta$
\footnote{We split the matrix $g_{AB}$ into two parts: its volume
and the distortion part, $
g_{AB}=e^{2\alpha}\left(e^{2\beta}\right)_{AB}. $ Where the scalar
$\alpha$ represents the volumetric expansion  whilst $\beta$ is
symmetric, trace free $3 \times 3$ matrix and hence $e^{2\beta}$ is
given by the series
$$e^{2\beta}=\sum_{r=0}^{\infty}{\frac{1}{r!}(2\beta)^r}.$$} are given by \citep{Misner}, 
\begin{equation} \label{alpha}
e^{\alpha}=\frac{h^{1/2}\Omega_0}{H_0(1-\Omega_0)^{3/2}}
\sinh^2{\left(\frac{h^{1/2}\tau}{2}\right)}, \,\,\,\,\,\, \beta=0.
\end{equation}

\begin{figure}[t]
 \begin{center}
  \leavevmode
 \includegraphics[scale=1, angle=0]{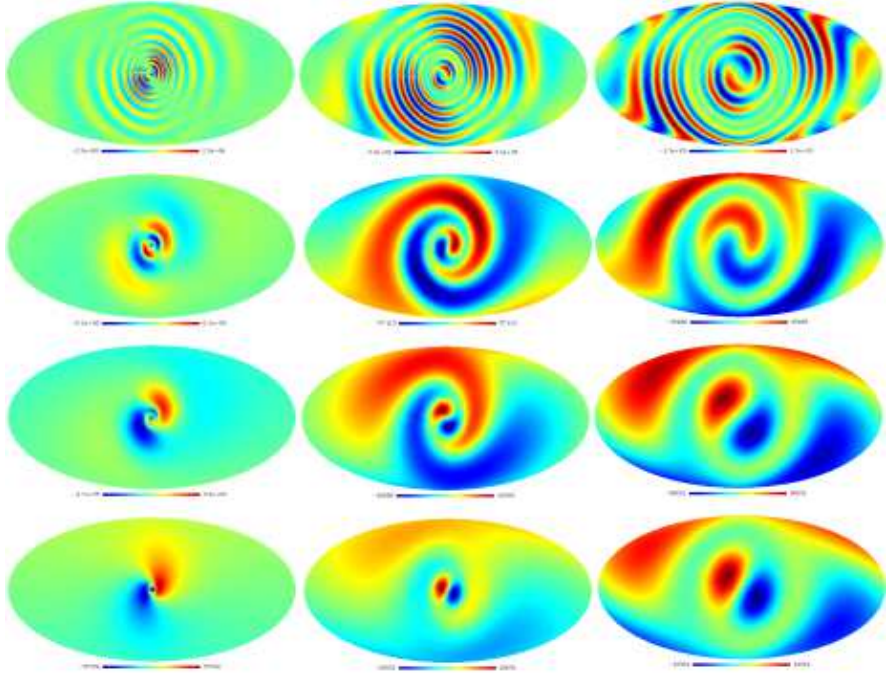}
  \caption[Bianchi VII$_h$ patterns for $x=5$]{Bianchi patterns in type VII$_h$ models for $x=0.1,\, 0.5,\, 1,\, 5$ (from top to bottom) and $\Omega_0=0.1,\,0.5$ and $0.9$ (from left to right). In low density models, the pattern is mostly focused in one hemisphere but for nearly flat models the spiral patterns widen and extend to both hemispheres. }%
  \label{bianchi4}
 \end{center}
\end{figure}

We introduce a factor $x=H_0e^{\alpha_0}$ which is related to the
parameter $h$ by 
\be 
x=\sqrt{\frac{h}{1-\Omega_0}}.
 \ee 
As we said
before, $x$ has no physical effect on the FRW models. It can be seen
from the fact that the present value of the scale factor can be
arbitrarily chosen, but from eqn. (\ref{alpha})  we have 
\be
e^{\alpha_0}= \frac{x}{H_0},
 \ee 
and hence, we see that parameter $x$
can be scaled out of the solution. However, for large values of $x$
(or equally $h$), the models are similar to those of type V. As the
present density tends towards the  critical density,
$\Omega_0\rightarrow 1$, and $h\rightarrow 0$ in such a way that $x$
remains finite,
 the behavior of the models tend to that of type VII$_0$. Behavior of these models
  with different parameters can be seen in figure \ref{bianchi4}.

Null geodesics have a complicated form and are given by
\begin{eqnarray}\label{nullgeodesicVII_h}
 \tan{\left(\frac{\theta}{2}\right)}&=&\tan{\left(\frac{\theta_0}{2}\right)}\exp{[-(\tau-\tau_0)\sqrt{h}]}
\\ \nonumber
\phi&=&\phi_0+(\tau-\tau_0)-\frac{1}{\sqrt{h}}\ln{\{\sin^2{\left(\frac{\theta_0}{2}\right)}+\cos^2{\left(\frac{\theta_0}{2}\right)}\exp{[2(\tau-\tau_0)\sqrt{h}]}\}}.
\end{eqnarray}
 They start
near the South Pole and spiral in the negative $\phi$-direction up
towards the equator, and then spiral in the positive
$\phi$-direction up towards the North Pole.  Small anisotropies can
be treated as perturbations to FRW model. The value of $\beta$ has been
calculated for them and is given in \cite{collins&hawking2}. Here we
follow \cite{barrow} who used these to calculate the contribution to
temperature anisotropies from the vorticity alone. They argue that
the inclusion of other pure shear distortions which are independent
of the rotation  could only make the temperature anisotropies
larger. And hence their results would give the maximum level of
vorticity consistent with a given value of temperature anisotropy.

In type VII$_h$, there are two independent vorticity components $\omega_2$ and $\omega_3$. They are given in terms of the off-diagonal shear elements as
\bea
\omega_2&=&\frac{(3h-1)\sigma_{13}-4h^{1/2}\sigma_{12}}{3x^2\Omega_0},
\\ \nonumber
\omega_3&=&\frac{(1-3h)\sigma_{12}-4h^{1/2}\sigma_{13}}{3x^2\Omega_0}.
\eea
The observables in this model are two dimensionless amplitudes: $\left(\frac{\sigma_{12}}{H}\right)_0$ and $\left(\frac{\sigma_{13}}{H}\right)_0$. The vorticity is given by
\be
\omega=\frac{1}{2}e^{-\alpha}(1+h)^{1/2}\left[(u_2)^2+(u_3)^2\right]^{1/2},
\ee
where $u_2$ and $u_3$ are  the velocity components and their present values are given by
\bea
\label{velocities}
(u_2)_0=\frac{1}{3x\Omega_0}\left[3h^{1/2}\left(\frac{\sigma_{12}}{H}\right)_0-\left(\frac{\sigma_{13}}{H}\right)_0\right],
\\ \nonumber
(u_3)_0=\frac{1}{3x\Omega_0}\left[\left(\frac{\sigma_{12}}{H}\right)_0+3h^{1/2}\left(\frac{\sigma_{13}}{H}\right)_0\right].
\eea
Hence the present value of vorticity is
\be
\left(\frac{\omega}{H}\right)_0=\frac{(1+h)^{1/2}(1+9h)^{1/2}}{6x^2\Omega_0}\left[\left(\frac{\sigma_{12}}{H}\right)_0^2+\left(\frac{\sigma_{13}}{H}\right)_0^2\right]^{1/2}.
\ee
To first order, temperature fluctuations are given by
\be
\label{DeltaTBianchi}
\frac{\Delta T(\theta_0,\phi_0)}{T_0}\simeq(p^iU_i)_0-(p^iU_i)_E-\int_E^0p^ip^k\sigma_{ik}\mathrm{dt},
\ee
where  $\theta_0$ and $\phi_0$ are related to the actual observing angles by
\be
\theta=\pi-\theta_0,
\,\,\,\,\,\,\,\,
\phi=\pi+\phi_0
\ee
Substituting for null geodesics from eqn. (\ref{nullgeodesicVII_h}) and for velocities from eqn. (\ref{velocities}) into eqn. (\ref{DeltaTBianchi}), we will obtain
\bea
\label{DeltaBianchiDerived}
\frac{\Delta T(\theta_0,\phi_0)}{T_0}&=&\left[\left(\frac{\sigma_{12}}{H}\right)_0A(\theta_0)+\left(\frac{\sigma_{13}}{H}\right)_0B(\theta_0)\right]\sin{\phi_0}
\\ \nonumber
&+& \left[\left(\frac{\sigma_{12}}{H}\right)_0B(\theta_0)-\left(\frac{\sigma_{13}}{H}\right)_0A(\theta_0)\right]\cos{\phi_0},
\eea
where coefficients $A(\theta_0)$ and $B(\theta_0)$ are defined by
\bea
\label{AandB}
A(\theta_0)&=&C_1[\sin{\theta_0}-C_2(\cos{\psi_E}-3h^{1/2}\sin{\psi_E})]
\\ \nonumber
&+&C_3\int_{\tau_E}^{\tau_0}\frac{s(1-s^2)\sin{\phi}\mathrm{d}\tau}{(1+s^2)^2\sinh^4{h^{1/2}\tau/2}},
\\ \nonumber
B(\theta_0)&=&C_1[3h^{1/2}\sin{\theta_0}-C_2(\sin{\psi_E}+3h^{1/2}\cos{\psi_E})]
\\ \nonumber
&-&C_3\int_{\tau_E}^{\tau_0}\frac{s(1-s^2)\cos{\phi}\mathrm{d}\tau}{(1+s^2)^2\sinh^4{h^{1/2}\tau/2}}.
\eea
The limits of integration are defined as
\bea
\tau_0=2h^{-1/2}\sinh^{-1}{(\Omega_0^{-1}-1)^{1/2}},
\\ \nonumber
\tau_E=2h^{-1/2}\sinh^{-1}{\left(\frac{\Omega_0^{-1}-1}{1+z_E}\right)^{1/2}},
\eea
and constants $C_1$, $C_2$, $C_3$, $s$ and $\psi$ are defined by
\bea
C_1&=&(3\Omega_0 x)^{-1};
\\ \nonumber
C_2&=&\frac{2s_E(1+z_E)}{1+s_E^2};
\\ \nonumber
C_3&=&4h^{1/2}(1-\Omega_0)^{3/2}\Omega_0^{-2};
\\ \nonumber
s&=& \tan{\left(\frac{\theta}{2}\right)}=\tan{\left(\frac{\theta_0}{2}\right)}\exp{[-(\tau-\tau_0)\sqrt{h}]},
\\ \nonumber
\psi&=&\phi_0+(\tau-\tau_0)-\frac{1}{\sqrt{h}}\ln{\{\sin^2{\left(\frac{\theta_0}{2}\right)}+\cos^2{\left(\frac{\theta_0}{2}\right)}\exp{[2(\tau-\tau_0)\sqrt{h}]}\}}.
\eea 
The expression for $\frac{\Delta T(\theta_0,\phi_0)}{T_0}$ can
be written in a compact form 
\be \label{compactform} \frac{\Delta
T}{T_0}(\theta_0,\phi_0)=(A^2+B^2)^{1/2}\left(\frac{\sigma}{H}\right)_0\cos{(\phi_0+\tilde{\phi})}
\ee 
where $\tilde{\phi}$ is defined as 
\be
\cos{\tilde{\phi}}=\left[\left(\frac{\sigma_{12}}{\sigma}\right)B-\left(\frac{\sigma_{13}}{\sigma}\right)A\right](A^2+B^2)^{-1/2}
\ee
 and $\sigma$ is given by 
\be
\sigma^2=\sigma_{12}^2+\sigma_{13}^2.
 \ee
 Eqn. (\ref{compactform})
helps to understand the behavior of the CMB pattern in these models.
If we look around any circle at a given $\theta_0$ on the sky, the
temperature variation will have a pure $\cos{\phi_0}$ behavior.
$\tilde{\phi}$ determines the relative orientation of adjacent
$\theta_0=$constant rings. On the other hand, $B(\theta_0)$
determines the amplitude of $\frac{\Delta T}{T_0}$ and the focusing
into a hot spot \citep{barrow}. Some of these maps have been shown
in figure \ref{bianchi4}.


\section{Unveiling Hidden Patterns of Bianchi VII$_h$}
\label{ourmethod}
 Recently Bianchi VII$_h$ models were compared to the first-year WMAP
 data on large scales and it was shown that the best fit Bianchi model
 corresponds to a highly hyperbolic model ($\Omega_0=0.5$) with a
 right-handed vorticity $(\frac{\omega}{H})_0=4.3 \times 10^{-10}$ \citep{jaffe}.
 There are several aspects to this hidden pattern which should be mentioned.
First of all, the proposed model is an extremely hyperbolic model and does not
agree with the location of the first peak in the best fit power spectrum of the
CMB. In the regime of  VII$_h$ models, one should study the nearly flat models in
order to have a consistent angular power spectrum. But the problem is that as it
has been shown in Fig. \ref{bianchi4}, the spiral patterns in high density models
are not concentrated in one hemisphere and can't be used to cure the large-scale power
asymmetry in the WMAP data reported by \cite{erik04a}.

Second is that the Bianchi patterns we studied so far, are only
valid for a matter dominated universe. Patterns must be recalculated
in a universe with a dark energy component if one wants to compare
them with the observed CMB data which is believed to be there in a
$\Lambda$ dominated universe\footnote{Bianchi-I patterns with dark
energy have been studied by \cite{tuhin}. Also \cite{Jaffe:2005gu}
have considered Bianchi VII$_h$ models with dark energy. }. These
issues are addressed in great details by \cite{Jaffe:2005gu} and
they conclude that the ``best-fit Bianchi type VII$_h$ model is not
compatible with measured cosmological parameters''.

For these reasons, we study the type VII$_h$ Bianchi models only as
hidden patterns in the CMB anisotropy maps. We pay our attention to
the specific model proposed by \cite{jaffe} and in the next two
sections, we address the question whether and to what extent one would
be able to discover this pattern and other hidden anisotropic patterns
in the CMB anisotropy maps. We carry out our analysis on full sky CMB
maps. It has been shown in our previous papers~\cite{us_bigpaper},
that for a masked CMB sky maps, the Bipolar power spectrum has a
calculable specific form. In that case, the violation of SI is
measured with respect to the `bias' that arises for the masked sky
map. The excess BiPS signature would certainly depend to some extent
on the specific orientation of the Bianchi template in the sky, given
by the extent to which the pattern is covered by the mask. In this
work we choose to keep our analysis independent of the orientation of
the hidden pattern.

\clearpage
\subsection*{Bipolar Power Spectrum Analysis}
We choose a Bianchi VII$_h$ model with
$\left(\frac{\sigma_{12}}{H}\right)_0=\left(\frac{\sigma_{13}}{H}\right)_0=\left(\frac{\sigma}{H}\right)_0$.
The temperature fluctuations induced by this model are given by eqn.
(\ref{DeltaBianchiDerived}) and will be \bea \frac{\Delta
T(\theta_0,\phi_0)}{T_0}= \left(\frac{\sigma}{H}\right)_0
[A(\theta_0)  &+& B(\theta_0)]\sin{\phi_0}
\\ \nonumber
+\,\, [B(\theta_0)&-&A(\theta_0)]\cos{\phi_0},
\eea
Power spectrum of the above Bianchi-induced fluctuations can be computed analytically \citep{barrow, McEwen} and is given by
\be
C_l = \frac{4\pi^2}{2l+1}\left(\frac{\sigma}{H}\right)_0^2 [(I_l^A)^2+(I_l^B)^2],
\ee
where
\bea
I_l^A &=& \sqrt{\frac{2l+1}{4\pi l (l+1)}}\int_0^\pi{A(\theta) P_l^1(\cos{\theta})\sin{\theta}\rm{d}\theta}
\\ \nonumber
I_l^B &=& \sqrt{\frac{2l+1}{4\pi l (l+1)}}\int_0^\pi{B(\theta)
P_l^1(\cos{\theta})\sin{\theta}\rm{d}\theta} \eea The majority of
the power in this Bianchi template is contained in multipoles below
$l\sim 20$. Therefor to study this particular model we do not need
high resolution CMB anisotropy maps.

\begin{figure}[h]
   \includegraphics[scale=0.8, angle=0]{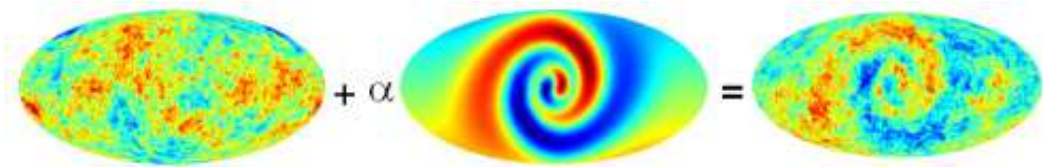}
   \caption[Adding a template to the background map]{Adding a pattern template with a strength $\alpha$ to a random realization of the CMB anisotropy map. This introduces a preferred direction in the map and violates the statistical isotropy. Maps are rotated to place the Galactic pole at the center for clear illustration.}
   \label{b+r}
 \end{figure}

To do a statistical study of the Bianchi patterns, we generate the
pattern for a given model. This pattern is given by the shear
components, $x$ parameter and the $\Omega_0$. We will then simulate
random CMB maps from the best fit $C_l$ of the WMAP data
\citep{sper_wmap03}. We add these two maps with a strength factor
$\alpha$ which will let us control the relative strength of the
pattern and the random map (see Fig. \ref{b+r}). The resultant map
is then given by
\be
\label{bianchideltat}
\Delta T(\hat{n}) \,=\,\Delta T^{CMB}(\hat{n})+\alpha \Delta T^{Bianchi}(\hat{n}),
\ee
and hence the power spectrum of this map will be given by
\be
\label{bianchicl}
C_l=C_l^{CMB}+\alpha^2 C_l^{Bianchi}
\ee
$\alpha$ is related to $(\frac{\sigma}{H})_0$ through $(\frac{\sigma}{H})_0$ =$\frac{\alpha}{T_0}$ where $T_0=2.73\times10^6\mu K$.
\begin{figure}[h]
  \includegraphics[scale=0.8, angle=0]{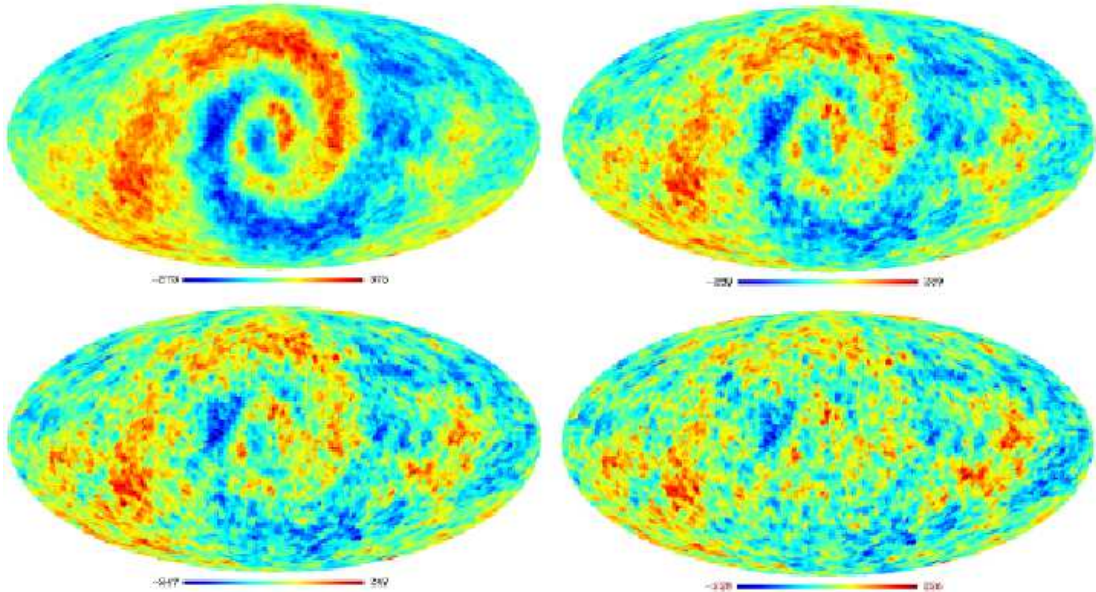}
  \caption[Bianchi added maps, $\Omega_0=0.5$]{Four Bianchi added CMB maps with different strength factors, rotated to place the Galactic pole at the center for clear illustration. The Bianchi template in the above maps has been computed with $x=0.55,\, \Omega_0=0.5$ and strength factors $(\frac{\sigma}{H})_0=1.83\times 10^{-9}$  ({\it top left}), $(\frac{\sigma}{H})_0=1.09\times 10^{-9}$ ({\it top right}), $(\frac{\sigma}{H})_0=7.3\times 10^{-10}$ ({\it bottom left}), $(\frac{\sigma}{H})_0=3.66\times 10^{-10}$ ({\it bottom right}). Although the pattern can hardly be seen in the fourth map, it has a considerable non-zero bipolar power spectrum (see Fig \ref{bips_bianchi_05}). }
  \label{bianchiaddedmaps_0.5}
\end{figure}

\begin{figure}[h]
  \includegraphics[scale=0.3, angle=-90]{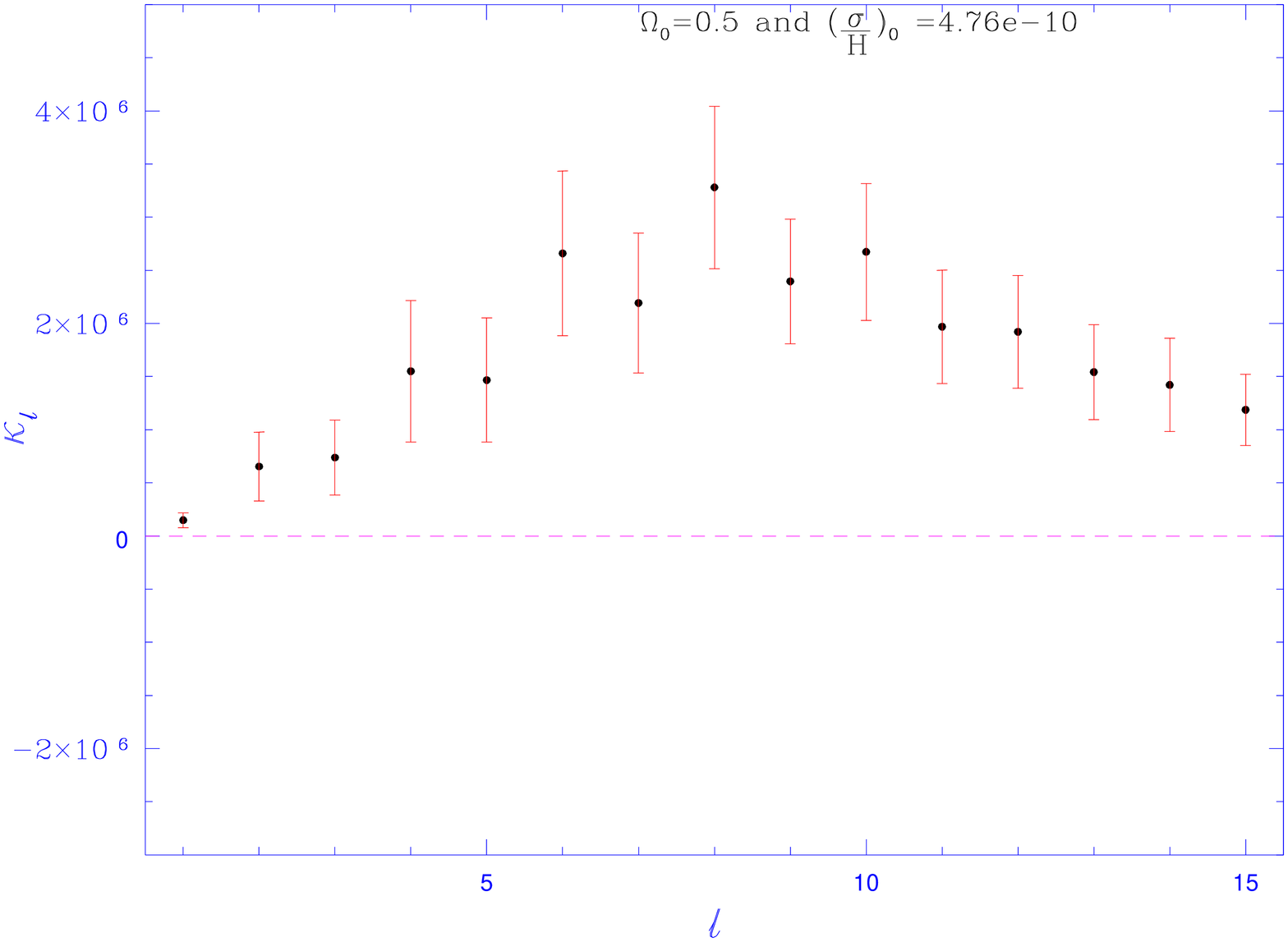}
\includegraphics[scale=0.3, angle=-90]{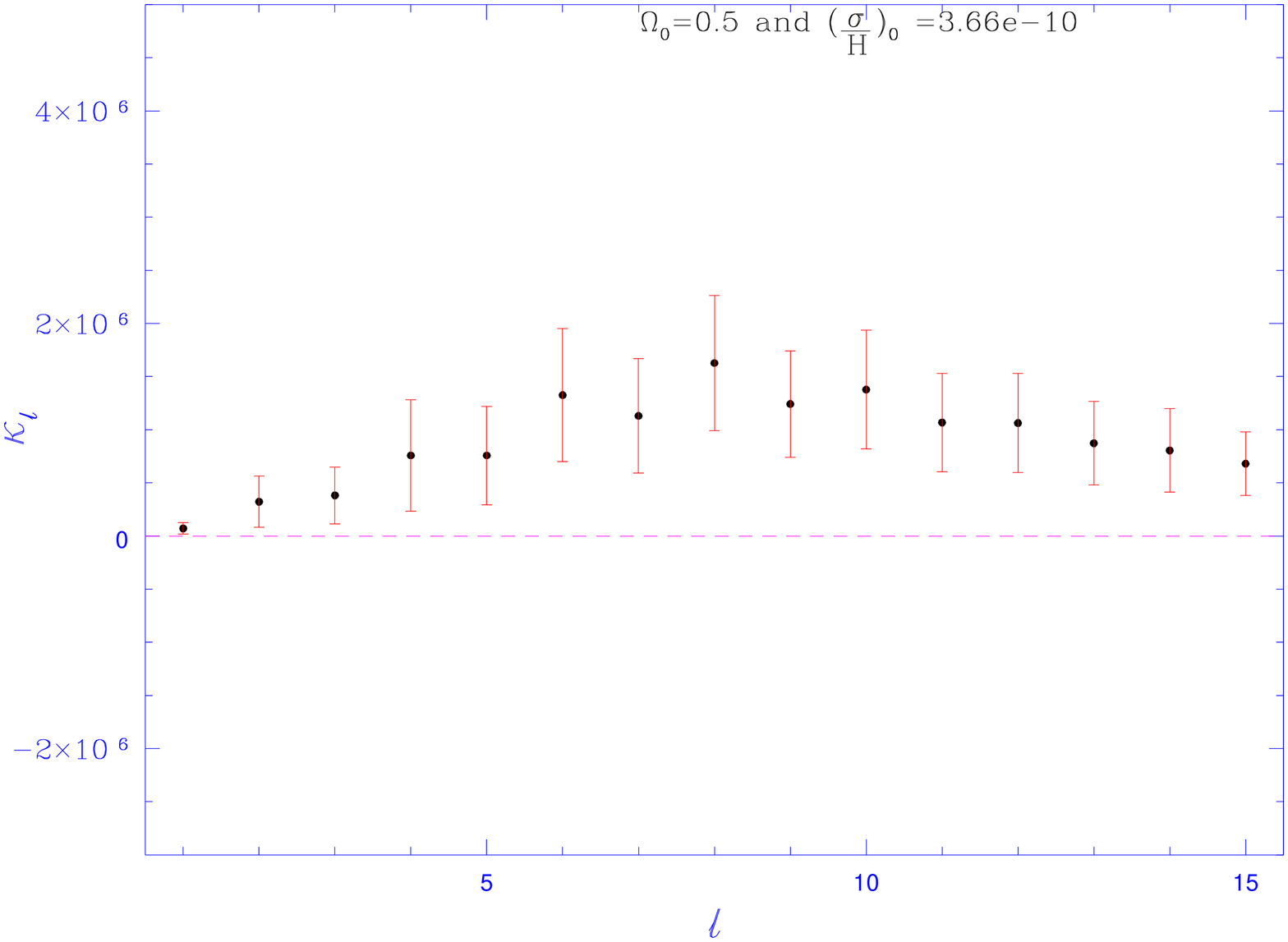}
\includegraphics[scale=0.3, angle=-90]{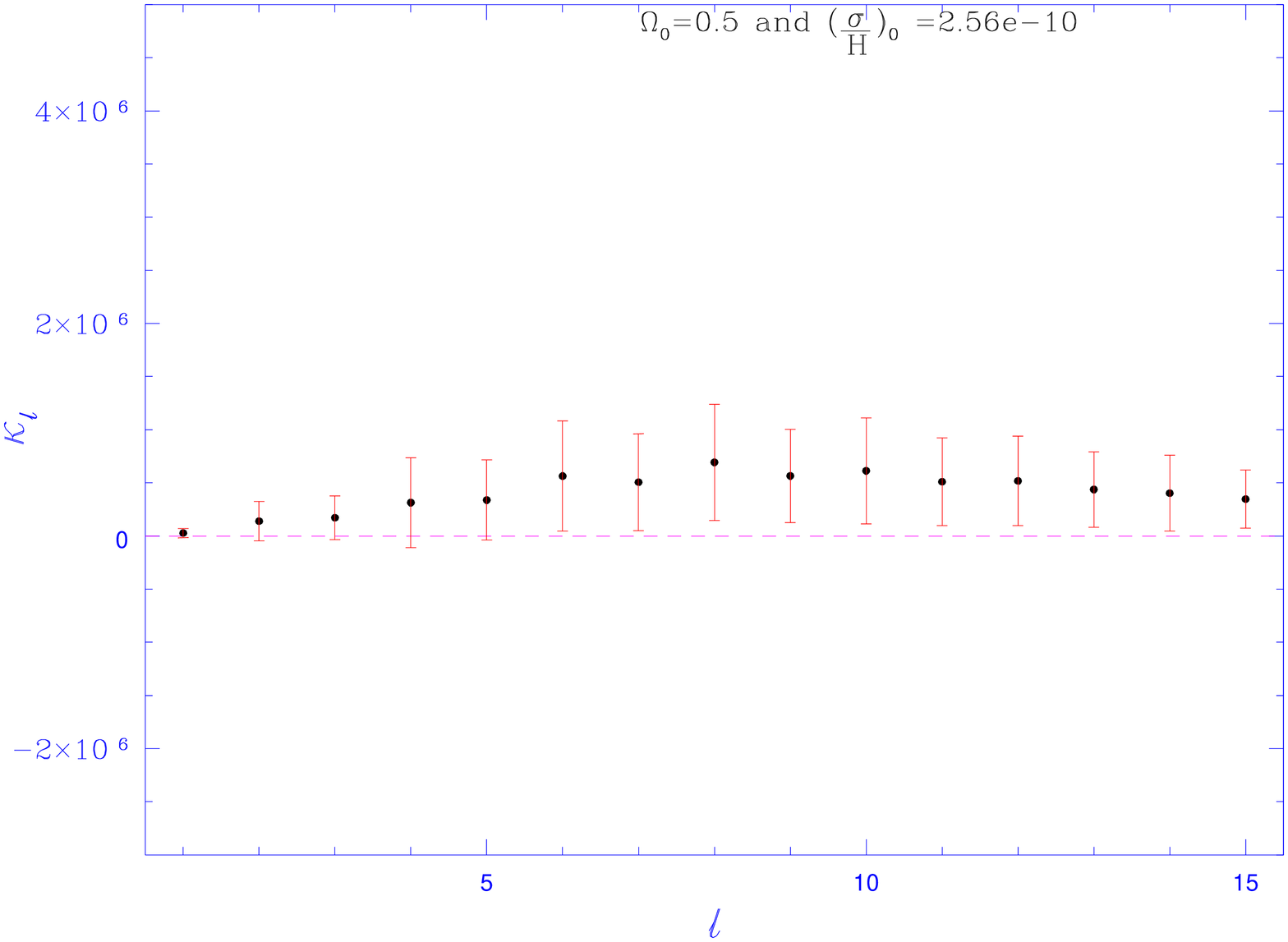}
\includegraphics[scale=0.3, angle=-90]{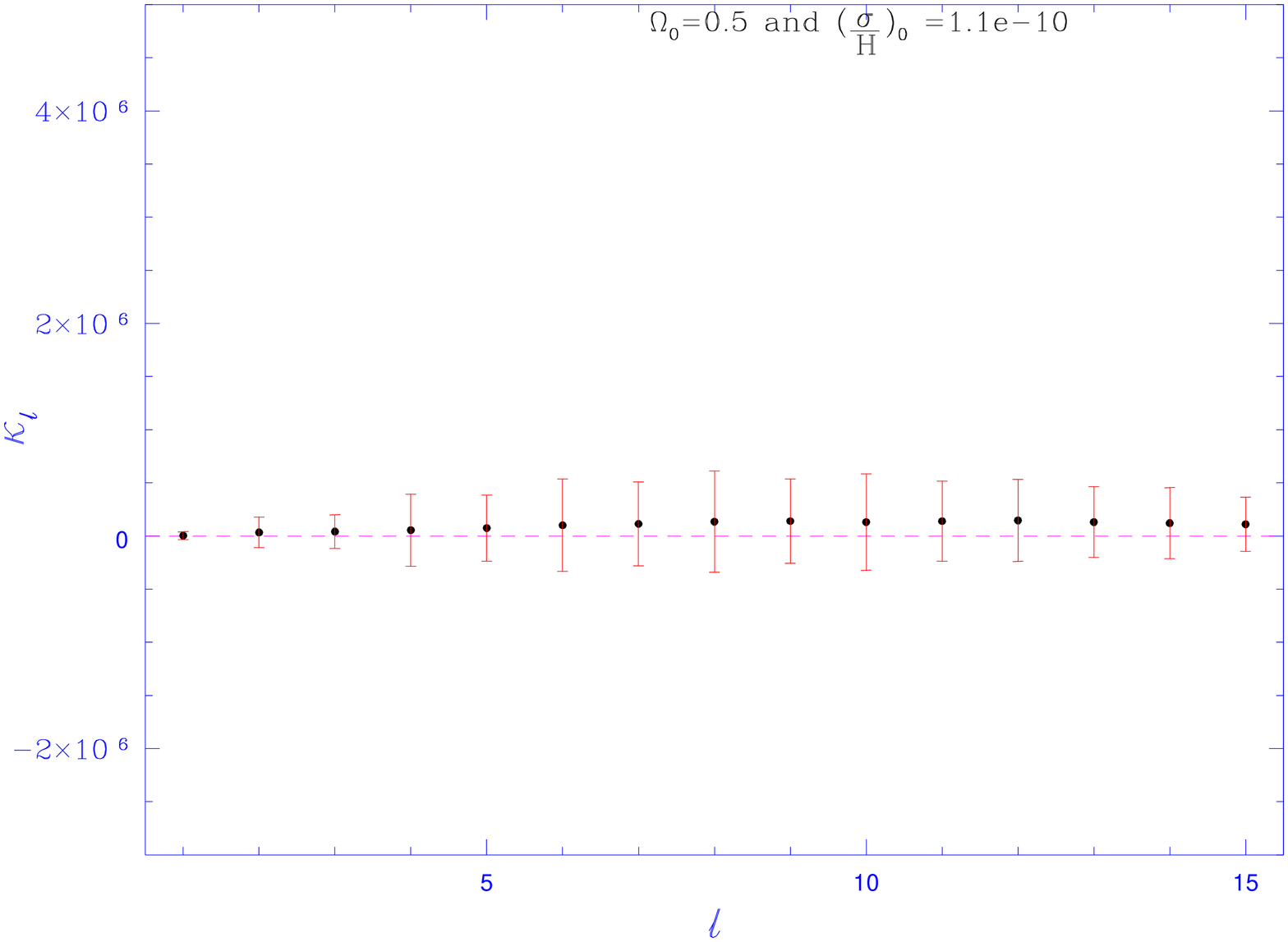}
  \caption[BiPS of Bianchi added maps, $\Omega_0=0.5$]{Bipolar power spectrum of Bianchi added CMB maps with strength factors of  $(\frac{\sigma}{H})_0=4.76\times10^{-10}, 3.66\times10^{-10}, 2.56\times10^{-10}, 1.25\times10^{-10}$. We have used $W_l^S(4,13)$ to filter the maps in order to focus on the low-$l$ (large angular scale) properties of the maps. }
  \label{bips_bianchi_05}
\end{figure}

We compute the bipolar power spectrum (BiPS) for the Bianchi added CMB
anisotropy maps generated above. The unbiased estimator of BiPS is
given by, \be \label{estimator}
\kappa_{\ell}=\sum_{ll'M}W_lW_{l'}|\sum_{mm'}a_{lm}a_{l'm'}\mathcal{C}^{\ell
M}_{ll'}|^2 - {\mathcal{B_{\ell}}}, \ee where $W_l$ is the window
function in harmonic space, ${\mathcal{B_{\ell}}}$ is the bias that
arises from the SI part of the map and is given by the angular power
spectrum, $C_l$ and $\mathcal{C}^{\ell M}_{ll'}$ are Clebsch-Gordan
coefficients. Bipolar power spectrum is zero for statistically
isotropic maps and has been studied in great details in
\cite{us_bigpaper} and \cite{us_polarization}. Although BiPS is
quartic in $a_{lm}$, it is designed to detect SI violation and not
non-Gaussianity \citep{us_apjl, us_bigpaper, us_bigpaper2}. Since BiPS
is orientation independent \citep{us_bigpaper}, we do not need to
worry about the relative orientation between the background map and
the Bianchi template.  For different strength factors we simulate 100
Bianchi added CMB maps\footnote{Simulation of background maps and the
spherical harmonic expansion of the resultant map is done using
HEALPix package \cite{healpix}.} at HEALPix resolution of
$N_{side}=32$ corresponding to $l_{max}=95$.  Some of these maps are
shown in Figure \ref{bianchiaddedmaps_0.5}. BiPS of each map is
obtained from the total $a_{lm}$ which according to eqn.
(\ref{bianchideltat}) are $a_{lm}=a^{CMB}_{lm}+\alpha \,
a^{Bianchi}_{lm}$. We compute the BiPS for each map using the
estimator given in eqn. (\ref{estimator}). We average these 100 BiPS
and compute the dispersion in them. The dispersion is an estimate of
$1\sigma$ error bars. At last we use the total angular power spectrum
given in eqn. (\ref{bianchicl}) to estimate the bias. We use the best
fit theoretical power spectrum from the WMAP analysis,
\cite{sper_wmap03}, as $C_l^{CMB}$ in computing the bias.  Some
results of BiPS for different strength factors are shown in Figure
\ref{bips_bianchi_05}. As it has been discussed in \cite{us_apjl,
us_bigpaper}, we can use different window functions in harmonic space,
$W_l$, in order to concentrate on a particular $l$-range.  This is
proved to be a useful and strong tool to detect deviations from SI
using BiPS method.  In other words, multipole space windows that weigh
down the contribution from the SI region of the multipole space will
enhance the signal relative to the cosmic error. We use simple filter
functions in $l$ space to isolate different ranges of angular scales;
low pass, Gaussian filters 
\be W_{l}^{G} = N^{G}
\exp\left\{-\left(\frac{2l + 1}{2l_{s} + 1} \right)^{2}\right\} \ee
that cut power on scales $(l \ge l_{s})$ and band pass filters of the
form \be W_{l}^{S} = 2 N^{S} \left[ 1 - J_{0}\left(\frac{2l + 1}
{2l_{s} + 1} \right)\right]\exp\left\{-\left(\frac{2l + 1}{2l_{s} + 1}
\right)^{2}\right\} ,
\ee where $J_0$ is the ordinary Bessel function
and $N^{G}$ and $N^{S}$ are normalization constants chosen such that,
\be \sum_{l} \frac{(2l + 1)W_{l}}{2l(l + 1)} = 1 \ee i.e., unit {\it
rms} for unit flat band angular power spectrum $C_{l} = \frac{2\pi}
{l(l + 1)}$.

We use a simple $\chi^2$ statistics to compare our BiPS results with zero,
\be
\chi^2 = \sum_{l=0}^{l_{max}} \left(\frac{\kappa_l}{\sigma_{\kappa_l}}\right)^2.
\ee
\begin{figure}[h]
\includegraphics[scale=1, angle=0]{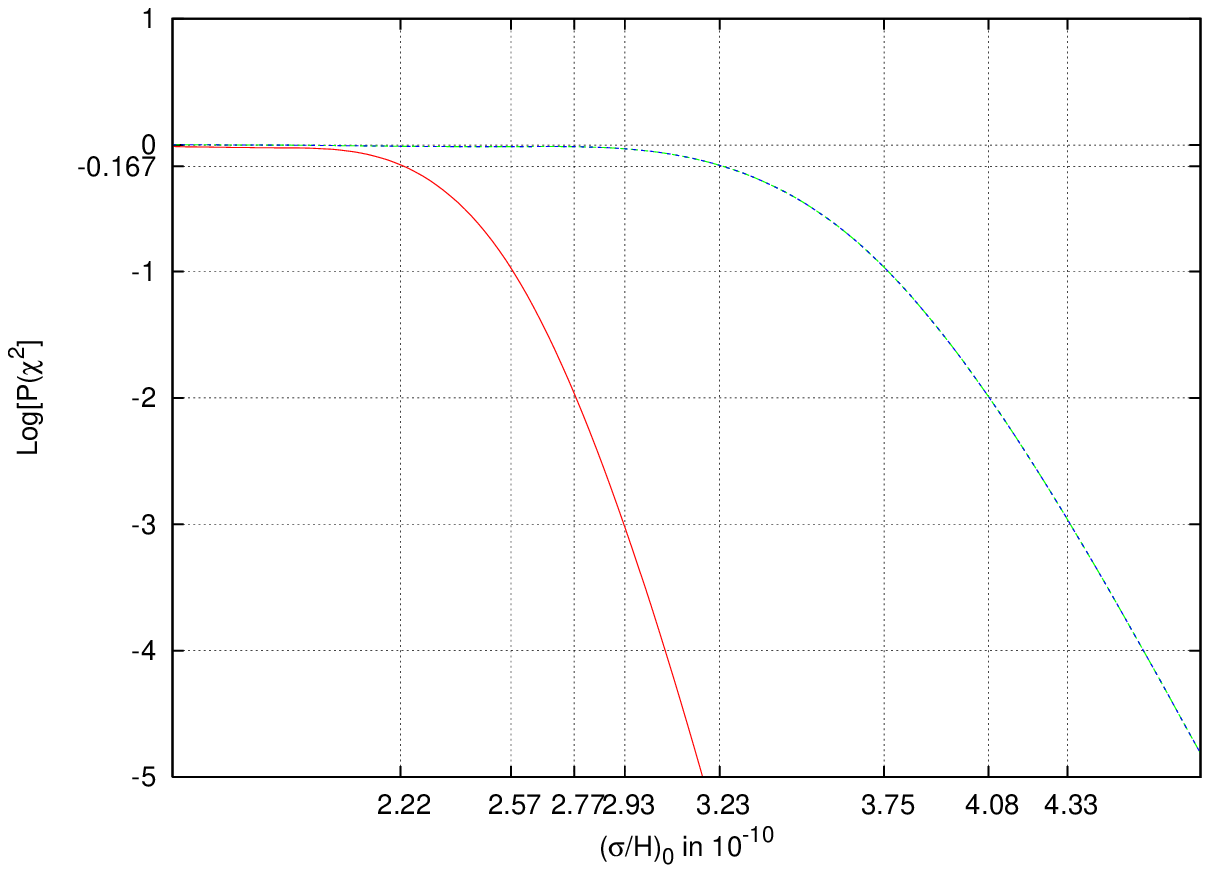}
\caption[Probability distribution of $\chi^2$ for low density Bianchi
template]{Probability distribution of $\chi^2$ versus the shear for
$\Omega_0=0.5$ Bianchi template. Results of two different window
functions, $W_l^G(25)$ (blue line) and $W_l^S(4,13)$ (red, solid line)
are shown.  By testing statistical isotropy of Bianchi embedded CMB
maps, we can put a limit of $(\frac{\sigma}{H})_0\le 2.77 \times
10^{-10} (99\% CL)$ on Bianchi VII$_h$ models.}
\label{probdist_bianchi_05}
\end{figure}
The probability of detecting a map with a given BiPS is then given
by the probability distribution of the above $\chi^2$. Probability
of $\chi^2$ versus the shear factor is plotted in Figure~\ref{probdist_bianchi_05}. Being on the conservative side we can
constrain the shear to be $(\frac{\sigma}{H})_0<
2.77\times 10^{-10}$ at a 99\% confidence level.

\section{ Conclusion}
Bianchi templates, the characteristic temperature patterns in CMB, are
imprinted by the unperturbed anisotropic expansion and have preferred
directions which violate statistical isotropy of the CMB anisotropy
maps. In this paper we study the consistency of the existence of a
hidden Bianchi template in the WMAP data that has proposed in recent
literature with the previous detection of null bipolar power spectrum
in the WMAP first year maps \citep{us_prd}.  We compute the bipolar
power spectrum for low density Bianchi VII$_h$ models embedded in
background CMB anisotropy maps with the power spectrum that best fits
the first year data of WMAP. We find non-zero bipolar power spectrum
for models with $(\frac{\sigma}{H})_0 \le 2.77\times 10^{-10}$. This
is inconsistent with the null bipolar power spectrum of the WMAP data
at $(99\% CL)$.  We conclude that correcting the WMAP first year maps
for the Bianchi template may make some anomalies in the WMAP data
vanish but this will be done at the expense of introducing other
anomalies such as preferred directions and violation of statistical
isotropy into the Bianchi corrected maps.

\begin{acknowledgments}
AH wishes to thank Lyman Page and David Spergel for helpful comments
on the manuscript. AH also thanks Pedro Ferreira, Joe Silk and Dmitry
Pogosyan for useful discussions. TS acknowledges useful discussions
with Anthony Banday and Kris Gorski. TG thanks IUCAA for the vacation
students program where the work was initiated. TG also thanks Somnath
Bharadwaj for help and co-guidance for the Master thesis at IIT,
Kharagpur.  AH acknowledges support from NASA grant
LTSA03-0000-0090. The computations were performed on Hercules, the
high performance computing facility of IUCAA. Some of the results in
this paper have used the HEALPix package. We acknowledge the use of
the Legacy Archive for Microwave Background Data Analysis
(LAMBDA). Support for LAMBDA is provided by the NASA Office of Space
Science.
\end{acknowledgments}

\appendix

\end{document}